\documentclass[12pt]{article}

\usepackage{amsmath,graphicx}
\usepackage{epsf}
\usepackage{graphicx,epsfig}
\usepackage{amsfonts}
\usepackage{amssymb}
\usepackage[nosort]{cite}
\usepackage{setspace}

\def\bk{{\bf k}}

\def\bx{{\bf x}}

\def\CL{{\cal L}}
\def\CO{{\cal O}}

\def\mpl{M_{\rm P}}
\def\half{\frac{1}{2}}

\makeatletter
\renewcommand\section{\@startsection {section}{1}{\z@}%
                                 {-3.5ex \@plus -1ex \@minus -.2ex}%
                                   {2.3ex \@plus.2ex}%
                                   {\normalfont\large\bfseries}}
\renewcommand\subsection{\@startsection{subsection}{2}{\z@}%
                                   {-3.25ex\@plus -1ex \@minus -.2ex}%
                                     {1.5ex \@plus .2ex}%
                                     {\normalfont\bfseries}}
\renewcommand\subsubsection{\@startsection{subsubsection}{3}{\z@}%
                                   {-3.25ex\@plus -1ex \@minus -.2ex}%
                                     {1.5ex \@plus .2ex}%
                                     {\normalfont\itshape}}
\makeatother

\newcommand{\Letter}{
\setlength{\textwidth}{16.5cm}
   \setlength{\textheight}{22.6cm}
    \hoffset=-0.5in
\voffset=-2.1cm }

\Letter

\renewcommand{\theequation}{\thesection.\arabic{equation}}

\begin{document}
\newcommand{\be}{\begin{equation}}
\newcommand{\ee}{\end{equation}}
\newcommand{\bea}{\begin{eqnarray}}
\newcommand{\eea}{\end{eqnarray}}
\newcommand{\barr}{\begin{array}}
\newcommand{\earr}{\end{array}}

\renewcommand{\theequation}{\arabic{equation}}

\thispagestyle{empty}
\begin{flushright}
\end{flushright}

\vspace*{0.3in}
\begin{spacing}{1.1}

\begin{center}
{\large \bf Fingerprints of Primordial Universe Paradigms
\\ \medskip as Features in Density Perturbations}

\vspace*{0.5in} {Xingang Chen}
\\[.3in]
{\em Center for Theoretical Cosmology, \\
Department of Applied Mathematics and Theoretical Physics, \\
University of Cambridge, Cambridge CB3 0WA, UK} \\[0.3in]
\end{center}

\begin{center}
{\bf
Abstract}
\end{center}
\noindent
Experimentally distinguishing different primordial universe paradigms that lead to the Big Bang model is an outstanding challenge in modern cosmology and astrophysics. We show that a generic type of signals that exist in primordial universe models can be used for such purpose. These signals are induced by tiny oscillations of massive fields and manifest as features in primordial density perturbations. They are capable of recording the time-dependence of the scale factor of the primordial universe, and therefore provide direct evidence for specific paradigm.

\vfill

\newpage
\setcounter{page}{1}


\newpage


Current observations have mapped out a detailed 13.7 billion years of Big Bang expansion history for our universe. One important knowledge we learned from these observations is that, at the very beginning of the Big Bang, there are tiny fluctuations in an otherwise very homogeneous and isotropic background density. These fluctuations, called the primordial density perturbations, become the seeds for the subsequent evolution of the large scale structures. They turn out to have very special properties. They are seeded at super-horizon scales, and are approximately scale-invariant, Gaussian and adiabatic \cite{Komatsu:2010fb}. Understanding the origin of these initial conditions and hence the origin of the Big Bang becomes an outstanding challenge for the modern cosmology.

The inflation paradigm \cite{Guth:1980zm,Linde:1981mu,Albrecht:1982wi} is the leading candidate for creating such a primordial universe. The simplest inflationary scenarios not only explain why our universe is homogeneous and isotropic, but also the properties of the density perturbations. Nonetheless, based on current data, there exist important ambiguities and degeneracies. Competing paradigms have been proposed to produce the same initial conditions. These alternatives include the paradigms of cyclic universe \cite{Khoury:2001wf}, matter bounce \cite{Wands:1998yp,Finelli:2001sr} and string gas \cite{Brandenberger:1988aj}. Arguably, the inflation still remains as the best available paradigm, because its generic predictions naturally fit the data and its microscopic origin in terms of fundamental physics is promising. However, such opinions may be subject to personal tastes; they are model-dependent and may even evolve with time. Besides theoretical establishment for each paradigm, a universal and unambiguous distinguisher is in terms of data. Namely we need to find some properties in the density perturbations that can clearly distinguish different paradigms. Such distinguishers are desirable even if one believes in a paradigm. In order to do so, such properties should satisfy the following requirements. They have to be shared by all general models in one paradigm, not just by a small subset of models; and they have to be distinctive for different paradigms.

So far the primordial tensor mode is regarded as the only possible solution to this question. They originate from gravitational waves in the primordial universe \cite{Starobinsky:1979ty}, and may be observed in terms of polarizations in the cosmic microwave background (CMB) \cite{Kamionkowski:1996zd,Seljak:1996gy}. Inflation models indeed have general predictions on the tensor mode. Namely, the tensor modes are approximately scale-invariant with a red tilt; for some models they are observable. Some alternative paradigms predict non-observable tensor modes, such as the cyclic universe paradigm \cite{Khoury:2001wf}, or observable ones with blue-tilt, such as the string gas paradigm \cite{Brandenberger:2006xi}. However there exist important caveats. Firstly, even for inflation, tensor modes sourced by the primordial gravitational waves are not guaranteed to be observable.
While the best sensitivity for the tensor-to-scalar ratio achievable by experiments in the near future is $\Delta r \sim \CO(10^{-3})$, the inflation models predict anywhere between $r \sim \CO(10^{-1})$, for large field models, and $r\sim\CO(10^{-55})$, for small field models with TeV-scale reheating energy. Secondly, if we consider more general alternatives, scale-invariant and observable tensor modes are possible. The equation of motion obeyed by each polarization component of the tensor modes is the same as that by the massless scalar with unit sound speed. So the tensor modes can be scale-invariant even in non-inflationary spacetime, just as the scalar. They become observable if the Hubble parameter is large. The matter contraction paradigm is such an example \cite{Wands:1998yp,Finelli:2001sr}.

Therefore it is very important to look for complimentary properties in the density perturbations that can be used as a model-independent distinguisher between different paradigms. This is the main purpose of this paper.

Almost by definition, different primordial universe paradigms are distinctively characterized by different time-dependence in the scale factor $a(t)$. To be explicit, we approximate the arbitrary time-dependent scale factor as the general power-law,
\bea
a(t) = a(t_0) (t/t_0)^p ~.
\label{powerlaw}
\eea
The density perturbations are seeded at superhorizon scales, so we require that the quantum fluctuations exit the event horizon $|a\tau|$, where $\tau$ is the conformal time defined by $dt \equiv ad\tau$ and is related to $t$ by $ a\tau = t/(1-p)$.
To satisfy this requirement, for $p>1$ we need an expansion phase, so $t$ runs from $0$ to $+\infty$; for $0<p<1$ we need a contraction phase, so $t$ runs from $-\infty$ to $0$; for $p<0$, we again need an expansion phase, so $t$ runs from $-\infty$ to $0$.
For example, $p>1$ corresponds to the inflation, $p=2/3$ the matter contraction, $0<p\ll 1$ the ekpyrotic (slowly contracting) phase, and $-1 \ll p <0$ the slowly expanding phase.
$\tau$ always runs from $-\infty$ to $0$.

Many properties of the density perturbations are convoluted consequence of this scale factor, and this is a primary reason for the ambiguity. Perhaps the most direct way to distinguish different paradigms would be to find some properties that can directly record the time-dependence of the scale factor. In this paper, we show that a type of feature models that generically exist in the primordial universe paradigms provide such opportunities.

We first describe this idea heuristically. Consider models with some small and repeated features. Such features can be various kinds of small structures in the realistic primordial universe models, and we will give a very general example shortly. These features generate small and repeated beats in cosmological parameters during the evolution, which can be used as a clock and related to the time $t$. In the mean while, the scale factor $a$ and the comoving momenta $k$ of quantum fluctuations are related by the physical momenta by definition.
The way that the features imprint themselves on the density perturbations is through the beats on the physical modes. Therefore the scale factor $a$ as a function of $t$ is translated to the comoving momenta $k$ as a function of features. What we can measure is features as a function of $k$. Knowing that we know the inverse function of $a(t)$. Let us now examine how this idea may be implemented and how general it is.

In order to achieve the goal, we need to look for standard physical clocks. Such clocks should generate repeated features with known time-dependence, although not necessarily periodic. They should exist generically so the signals are more likely to be observed, and they should also be associated with a set of specific patterns identifiable in observations. The classical oscillation of massive fields are ideal for such purposes.

Massive fields with mass much larger than the horizon mass-scale $1/|a\tau|$ exist ubiquitously in models of primordial universe. Even when we think of single field models, in the context of a unification theory, what we really have in mind is models with many massive modes. The single field model is obtained as the low energy limit at or below the energy scale $1/|a\tau|$, after the massive modes are integrated out. If we take into account these massive field directions, the trajectory of the effective single field $\phi$ in this multi-dimensional field space is unlikely a straight line. Generically it should turn from time to time. During the turning, the field $\phi$ and the massive modes couple, and this induces the energy transfer. The massive fields are excited and oscillate around the potential minima classically. Because generically the fraction of energy transfer is small, the oscillation amplitudes are small but the frequency are large. So for most purposes, these effect are averaged out and safely ignored. But for our purpose, these small oscillations are a good candidate for the physical clock that we are looking for. So let us examine more closely the behavior of the massive field $\sigma$ in the power-law background.

The equation of motion is
\bea
\ddot \sigma + 3H\dot\sigma + m_\sigma^2 \sigma =0 ~,
\eea
where the Hubble parameter $H=p/t$. The solution is given in terms of Bessel functions. The asymptotic behavior of these Bessel functions at the limit $m_\sigma t\gg p^2$ is given in terms of sinusoidal functions, and we use these to approximate the oscillatory behavior of $\sigma$,
\bea
\sigma \approx \sigma_A \left( \frac{t}{t_0} \right)^{-3p/2}
\left[ \sin(m_\sigma t +\alpha) + \frac{-6p+9p^2}{8m_\sigma t} \cos(m_\sigma t + \alpha) \right] ~,
\label{sigmaOsci}
\eea
where $\alpha$ is a phase, and $\sigma_A$ is the initial oscillation amplitude at $t=t_0$. Such oscillations induce an oscillatory component to the Hubble parameter $H$, because
\bea
3\mpl^2 H^2 = \half \dot\sigma^2 + \half m^2\sigma^2 + {\rm other~fields}~.
\label{Heom}
\eea
The leading term on the right hand side of (\ref{Heom}) does not oscillate in time because the energy is converting between kinetic and potential energy back and forth and conserved in the leading order. The oscillatory component for $H$, which we denote as $H_{\rm osci}$, comes from the subleading terms. Using (\ref{sigmaOsci}), we get
\bea
H_{\rm osci} = - \frac{\sigma_A^2 m_\sigma}{8\mpl^2} \left(\frac{t}{t_0}\right)^{-3p} \sin(2m_\sigma t + 2\alpha) ~.
\eea
This in turn induces the oscillatory components for the parameters $\epsilon \equiv -\dot H/H^2$ and $\eta \equiv \dot\epsilon/(H\epsilon)$. Again we use the subscript ``${\rm osci}$" to denote their oscillatory components,
\bea
\epsilon_{\rm osci} &=& \frac{\sigma_A^2 m_\sigma^2}{4 \mpl^2 H^2} \left( \frac{t}{t_0} \right)^{-3p} \cos(2m_\sigma t + 2\alpha) ~,
\label{epsilon_osci}
\\
\dot\eta_{\rm osci} &=& - \frac{\sigma_A^2 m_\sigma^4}{\mpl^2 \epsilon H^3} \left( \frac{t}{t_0} \right)^{-3p} \cos(2m_\sigma t + 2\alpha)
\label{doteta_osci}
~.
\eea

Generally, multiple massive fields are excited sporadically along the trajectory with different spectra. These oscillations induce corresponding beats in the cosmological parameters. The question now is how sensitive the observables are to these oscillations and how they can be identified in experiments. To study this, let us turn to the subject of Bunch-Davies (BD) vacuum and resonance mechanism.

In nearly all models of primordial universe, the quantum fluctuations start their life in a vacuum that is mostly BD. These fluctuations later exit the event horizon and become the seeds for the large scale structure.
Their general property can be seen as follows.
Consider the fluctuations of an effectively massless scalar field, $\delta\phi(\bx,t)$, in a general time-dependent background with the scale factor $a(t)$,
\bea
\CL= \frac{a^3}{2} (\dot {\delta\phi})^2 - \frac{a}{2} (\partial_i \delta\phi)^2 ~.
\label{Ldeltaphi}
\eea
A quantum fluctuation with comoving momentum $k$ is within the event horizon if $k>1/|\tau|$. In this limit, the equation of motion for the fluctuations approaches that in the Minkowski spacetime limit. Along with the quantization condition,
\bea
a^3 \delta\phi \dot{\delta\phi}^* - {\rm c.c.} = i ~,
\eea
the mode function in the subhorizon limit becomes
\bea
\delta\phi \to \frac{1}{a \sqrt{2k}} e^{-ik\tau} ~.
\label{BDphi}
\eea
We have chosen the positive-energy mode, which corresponds to the ground state of the Minkowski spacetime, to be the BD vacuum. The effect of the background time-dependence is incorporated adiabatically in (\ref{BDphi}). The most important and universal property of (\ref{BDphi}) is the oscillatory factor $e^{-ik\tau}$.

While the superhorizon evolution of different paradigms are very model-dependent, the subhorizon BD-vacuum behavior is universal. This applies to both inflationary and non-inflationary scenario, expansion and contraction universe, attractor and non-attractor evolution, single field and multifield model, and curvaton and isocurvaton modes.

If the background cosmological parameters have small oscillatory components with frequencies much larger than the energy scale of the event-horizon, they introduce a high energy probe to the BD vacuum by resonating with the vacuum component that has the same frequency \cite{Chen:2008wn}. Since the BD vacuum has the background time-dependence, different momentum modes get resonated at different time. This process produces large correlation amplitudes and distinctive scale-dependence (running). They manifest as features with specific patterns in density perturbations, which we will compute shortly.

To summarize, we see that three universal properties nicely fit into each other for our purpose. Firstly, the generic oscillation of fields with mass much larger than $1/|a\tau|$ can be used as the standard clock, and they generate small beats with known time-dependence in cosmological parameters. Secondly, these oscillations affect the density perturbations through the universal BD-vacuum instead of the model-dependent superhorizon behavior, which makes general analyses possible. Thirdly, the subhorizon scale is precisely where the strong resonance effect takes place, generating large effect on density perturbations.

We consider a two-field model in the general time-dependent background as the explicit example. The $\phi$ field is responsible for the leading order scale-invariant power spectrum $P_{\zeta 0}$. This field may be the one driving the background evolution, or a different field that is introduced to source the density perturbations only. The details are highly model-dependent, but this is not our concern here. We are interested in computing the resonance effects induced by massive fields as corrections to the leading power spectrum and as the leading bispectra. To be as general as possible, after the massive field $\sigma$ is excited and starts to oscillate at time $t_0$, we consider the massive field as a spectator which couples to $\phi$ only through gravity,
\bea
\CL = \sqrt{-g} \left[ -\half g^{\mu\nu} \partial_\mu \phi \partial_\nu \phi - V_\phi(\phi)
-\half g^{\mu\nu} \partial_\mu \sigma \partial_\nu \sigma - \half m_\sigma^2 \sigma^2 \right] ~.
\label{twofield_L2}
\eea
More complicated couplings are of course possible, but this is the minimum case. The correlation functions for the scalar perturbation $\zeta$ are computed in the in-in formalism. See Ref.~\cite{Chen:2011zf} for computational details and Ref.~\cite{Chen:2010xk} for general methods. For power spectrum, we have
\bea
\frac{\Delta P_\zeta}{P_{\zeta 0}} = 2i \int_{-\infty}^0 d\tau~
a^2 \epsilon_{\rm osci}~
({u_{k_1}'}^2 - k_1^2 u_{k_1}^2 ) + {\rm c.c.} ~.
\label{powerspectrum_corr}
\eea
For bispectrum, we have
\bea
\langle \zeta_{\bk_1} \zeta_{\bk_2} \zeta_{\bk_3} \rangle
&=& i \left( \prod_i u_{k_i}(0) \right)
\int_{-\infty}^0 d\tau~ a^3\epsilon_0 \dot\eta_{\rm osci} ~
u_{k_1}^* u_{k_2}^* \frac{du_{k_3}^*}{d\tau}
\nonumber \\
&\times& (2\pi)^3 \delta^3(\sum_i \bk_i) + {\rm 2~perm.} + {\rm c.c.} ~.
\label{bispectrum}
\eea
We will quote the bispectrum in terms of a function $S(k_1,k_2,k_3)$ defined as \cite{Chen:2010xk}
\bea
\langle \zeta_{\bk_1} \zeta_{\bk_2} \zeta_{\bk_3} \rangle =
S(k_1,k_2,k_3) \frac{1}{(k_1k_2k_3)^2} P_{\zeta 0}^2 (2\pi)^7 \delta^3(\sum_i \bk_i) ~.
\eea
The mode function $u_k(\tau)$ in these expressions is the Fourier transform of $\zeta(\bx, \tau)$, and as mentioned only the BD behavior is needed in our computation. In the above integrals, the resonance happens between the oscillatory components of the parameters, $\epsilon_{\rm osci}$ and $\dot\eta_{\rm osci}$, and the mode functions.

For arbitrary $p$, both power spectra and bispectra have similar $p$-dependent resonant running behavior,
\bea
\frac{\Delta P_\zeta}{P_{\zeta 0}} &\propto&
\left( \frac{2k_1}{k_r} \right)^{-3+\frac{5}{2p}}
\sin \left[ \frac{p^2}{1-p} \frac{2m_\sigma}{H_0} \left( \frac{2k_1}{k_r} \right)^{1/p} + {\rm phase} \right] ~,
\label{res_powerspectrum}
\\
S &\propto&
\left( \frac{K}{k_r} \right)^{-3+\frac{7}{2p}}
\sin \left[ \frac{p^2}{1-p} \frac{2m_\sigma}{H_0}
\left( \frac{K}{k_r} \right)^{1/p} + {\rm phase} \right] ~.
\label{res_bispectrum}
\eea
The resonant running refers to the scale-dependence of the functions $\sin[\dots]$ in these expressions.
We have also included the typical scale-dependence of the amplitudes in front of these functions. The parameter $k_r$ denotes the first resonating $K$-mode when the massive field starts to oscillate ($K=2k_1$ for power spectrum and $K=k_1+k_2+k_3$ for bispectrum). For the expanding background, $p>1$ and $p<0$, lower $K$-modes resonate earlier and the above formulae apply to $K>k_r$; for the contracting background, $0<p<1$, larger $k$-modes resonate earlier and they apply to $K<k_r$.

The most important property in the above results is that the resonant running for different paradigms are distinctive. The differences are clear even within a couple of efolds.
This demonstrates the idea of using features to probe the scale factor. The arguments of the sinusoidal functions are power-law functions of comoving momenta with a power $1/p$, which is the inverse function of the power-law in the scale factor.

Generally speaking the overall sizes of these signals depend on detailed models. If we restrict to the inflation case, $p\gg 1$, the full results become
\bea
\frac{\Delta P_\zeta}{P_{\zeta 0}} &=&
\frac{\sqrt{\pi}}{4} \frac{\sigma_A^2}{\epsilon_0 \mpl^2}
\left( \frac{m_\sigma}{H} \right)^{5/2}
\left( \frac{2k_1}{k_r} \right)^{-3}
\sin \left[ \frac{2m_\sigma}{H} \ln 2k_1 + \tilde \alpha \right] ~,
\label{Res_2pt_inf}
\\
S &=& - \frac{\sqrt{\pi}}{8}
\frac{\sigma_A^2}{\epsilon_0 \mpl^2}
\left( \frac{m_\sigma}{H} \right)^{9/2}
\left( \frac{K}{k_r} \right)^{-3}
\sin \left[ \frac{2m_\sigma}{H} \ln K + \tilde \alpha \right] ~,
\label{Res_3pt_inf}
\eea
where the phase $\tilde \alpha = (2m_\sigma/H)(1-\ln 2m_\sigma) +2\alpha+\pi/4$. We will use these to demonstrate the sizes of the signals in the inflation case. The resonant running in this limit reproduces the form found by Chen, Easther and Lim (CEL) for inflation with periodic features \cite{Chen:2008wn}. In retrospect, the fact that the arguments are proportional to $\ln K$ tells us that this is inflation.

At least for inflation models, the amplitudes of the resonant power spectra and bispectra, which we label as $(\Delta P_\zeta/P_{\zeta 0})_A$ and $f_{NL}$ respectively, can be easily made very large.
We denote the fraction of the kinetic energy of $\phi$, that is converted to the energy in the $\sigma$-field during the turning and induces its oscillation, as $\beta$. So $m_\sigma^2 \sigma_A^2 \sim \beta \dot\phi^2$. Also note $\epsilon_0 \sim \dot\phi^2/(\mpl^2 H^2)$. From (\ref{Res_2pt_inf}) and (\ref{Res_3pt_inf}), we have
\bea
\left( \frac{\Delta P_\zeta}{P_{\zeta 0}} \right)_A
\sim \beta \left( \frac{m_\sigma}{H} \right)^{1/2} ~,
\quad
f_{NL} \sim \beta \left( \frac{m_\sigma}{H} \right)^{5/2} ~.
\label{Est_Bispectrum}
\eea
Even for a tiny fraction of energy transfer, the resonance amplitudes can be quite large. For example, for $\beta \sim 10^{-2}$, $m_\sigma/H \sim 10^2$, we have $\Delta P_\zeta/P_{\zeta 0} \sim 0.1$ and $f_{NL} \sim 10^3$. The resonance mechanism plays a crucial role in producing the large amplitudes. With same $\beta$, increasing the mass $m_\sigma$ reduces the oscillation amplitude $\sigma_A$, but this is overcome by the strength of the resonance which is proportional to higher powers of $m_\sigma$.

It is important to note that the distinctive oscillatory running behavior in the above resonant forms will not be changed by curvaton-isocurvaton couplings in multifield evolution. Any effect that oscillates faster than the horizon time-scale generates additional resonance forms that superimpose onto each other. Any effect that varies much slower can only change the overall envelop of the resonance forms, by either changing their overall sizes or introducing scale-dependent modulations. This is another crucial point that makes this type of features a model-independent general distinguisher between paradigms.

There are also interesting scale-dependence in the amplitudes. These dependence is much milder in the sense that the range of scales $\Delta K$ over which significant variation takes places is larger than or comparable to the local scale $K$ itself. For general multifield models, such scale dependence can be changed due to scale-dependence in curvaton-isocurvaton couplings. So they may not be as faithful as the resonant running to the specific paradigm. Nonetheless, as we will see, there are some very sharp differences between paradigms that can be used as supportive evidences.

It is also important to note the caveats in our arguments. In principle one may be able to engineer artificial features in a model that mimic the signals we get so far. The type of clocks we considered are periodic; but one may engineer non-periodic background oscillation that resonates in a non-inflationary background, and they conspire to give the same resonant running of the CEL form as in (\ref{Res_3pt_inf}). For instance, the oscillation of the form $\sim e^{ig\ln(t/t_0)}$ in the non-inflationary power-law background induces the same CEL form. Another possible caveat is that the mass may be time-dependent so the period of the clock changes.

These ambiguities can be reduced by looking for multiple characteristic signals. The main point is to strength the evidence that these features are induced by periodically oscillating massive fields.
We have so far only concentrated on the resonant running behavior. There are at least two types of multiple signals characteristic of massive fields.

\begin{figure}
\begin{center}
\epsfig{file=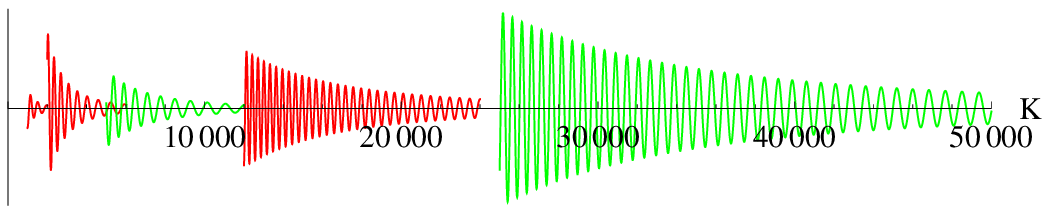, width=120mm}
\epsfig{file=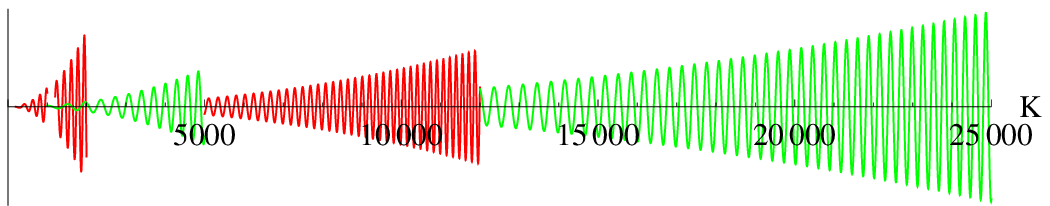, width=120mm}
\epsfig{file=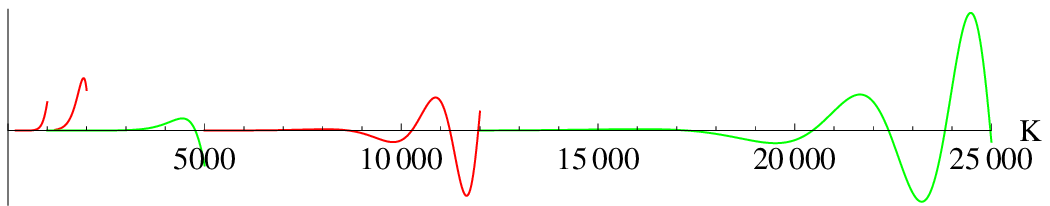, width=120mm}
\epsfig{file=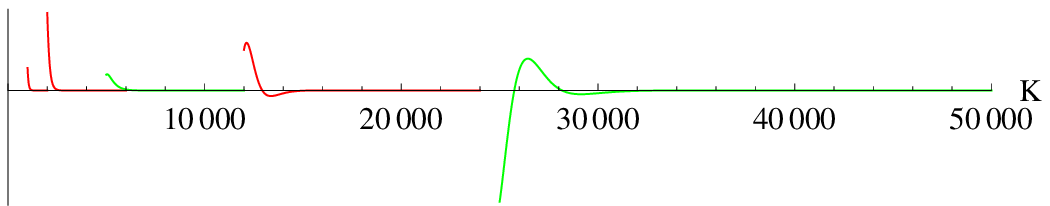, width=120mm}
\end{center}
\caption{The first type of fingerprints for different paradigms: multiple oscillations. From top to bottom: $p=10$ (inflation), $p=2/3$ (matter contraction), $p=0.2$ (Ekpyrosis), $p=-0.2$ (slow expansion). The red/dark spectra correspond to $m_\sigma/H = 10,20,120$ excited by a common sharp feature; the green/ligh spectra correspond to $m_\sigma/H=30,150$ excited by another common sharp feature.}
\label{Fig:multiple}
\end{figure}

Firstly, the density of massive fields is proportional to $a^{-3}$; in addition, they induce a constant resonance scale while the event-horizon sizes have different time-dependence for different paradigms. These properties determine the different running behavior of the amplitudes in (\ref{res_powerspectrum}) and (\ref{res_bispectrum}).
In particular, inflationary background has a very red running index $-3$, while the Ekpyrosis has a very blue running index $\propto 1/p$.
Although these types of running are less robust than the resonant running, this dramatic difference can be used as supportive evidence.
These running amplitudes will be modified if the mass is time-dependent. For example, for inflation, to make the resonant running deviate significantly from the CEL form, we need $\dot m_\sigma/(m_{\sigma}H) \gtrsim \CO(1)$; but this introduces dramatic scale-dependence in (\ref{Est_Bispectrum}).
In general, since the massive fields are excited sporadically at different places with different spectra, we expect that each paradigm has its unique fingerprints consisting of such multiple signals of the same type. This is illustrated in Fig.~\ref{Fig:multiple}.

\begin{figure}
\begin{center}
\epsfig{file=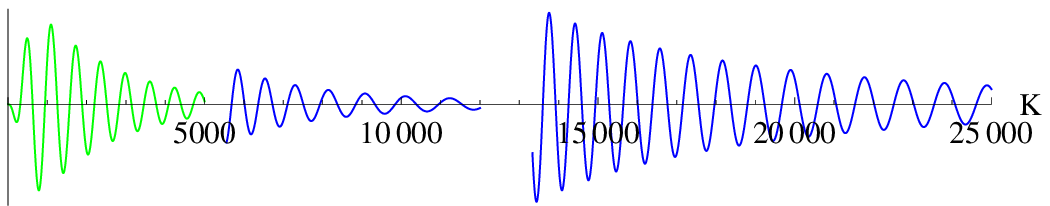, width=120mm}
\epsfig{file=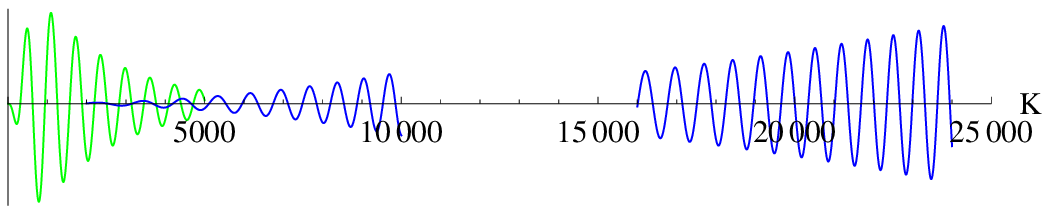, width=120mm}
\epsfig{file=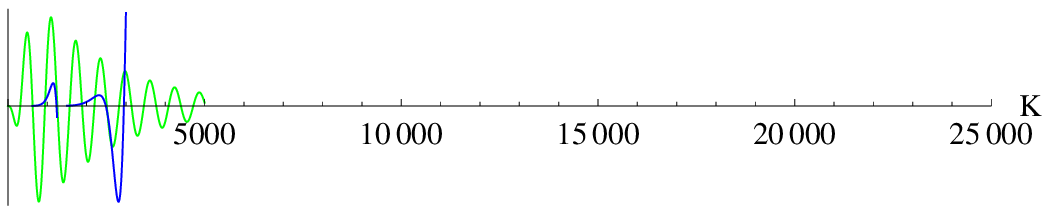, width=120mm}
\epsfig{file=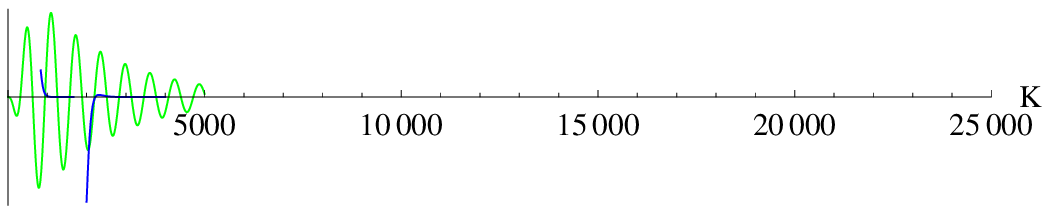, width=120mm}
\end{center}
\caption{The second type of fingerprints for different paradigms: in conjunction with sharp feature. From top to bottom: $p=10$ (inflation), $p=2/3$ (matter contraction), $p=0.2$ (Ekpyrosis), $p=-0.2$ (slow expansion). The Green/light spectra are generated by a sharp feature at $K=k_0=100$ and have sinusoidal running; the blue/dark spectra correspond to two massive fields ($m_\sigma= 25,60$) excited by this sharp feature and have resonant running.}
\label{Fig:sharp}
\end{figure}

Secondly, oscillations of the massive fields are triggered by some sharp physical processes, such as a sharp turning (with large angular velocity $\dot\theta\gg H$ but not necessary large turning angle). Such sharp features are also associated with characteristic observational signatures. In contrast to the resonance case, sharp feature signals are universal to different time-dependent backgrounds. See Ref.~\cite{Chen:2011zf} for details. The characteristic property is the sinusoidal running
\bea
\sim \sin (K/k_0 + {\rm phase}) ~,
\label{sin_running}
\eea
where $k_0$ is the starting point of this running and it is related to the conformal time $\tau_0$ at the location of the feature by $k_0\equiv |\tau_0|^{-1}$. Such signals cannot be used to distinguish different paradigms, but they can be used to identify the location of the sharp feature and the subsequent massive modes oscillations. Because of this connection, the parameters for these two different runnings, (\ref{res_bispectrum}) and (\ref{sin_running}), are related,
\bea
\frac{k_r}{k_0} = \frac{p}{|1-p|} \frac{2m_\sigma}{H_0} ~,
\label{krk0_relation}
\eea
where $H_0$ is the Hubble parameter evaluated at $\tau_0$. This is illustrated in Fig.~\ref{Fig:sharp}.

These signals present special opportunities and challenges for experiments and data analyses. Typically they are highly oscillatory in $K$-space, which means that in the multipole space the binning interval $\Delta \ell$ has to be much smaller than the local $\ell$, $\Delta\ell/\ell \sim H/m_\sigma \ll 1$. So experiments capable of measuring density perturbations in high multipoles are required. The Planck satellite is observing the CMB at multipoles of a few thousands. The ground-based telescopes, South Pole Telescope (SPT) \cite{Keisler:2011aw} and Atacama Cosmology Telescope (ACT) \cite{Hlozek:2011pc}, can go up to ten thousands. More speculatively, the 21cm hydrogen lines may be observed in much lower redshift and with much higher multipoles.

The CEL form, and most other resonance forms, have highly oscillatory and characteristic running behavior. This can easily make them orthogonal to many other primordial signals and late-time contaminants. Non-linear gravity effects and CMB evolution limit the detection sensitivity for conventional primordial bispectra to be $f_{NL}\sim 1$. For SPT and ACT, astrophysical effects such as point sources and Sunyaev-Zeldovich effect dominate over the primordial information for $\ell \gtrsim 2500$. However, since the resonant forms are orthogonal to these non-primordial contaminants, new assessment is necessary to find out the main limiting factors and make forecasts; and new observationally accessible windows can be opened up to probe the primordial universe even for existing experiments.

Conventional data analyses methods start with a specific separable template. But here a scan over families of non-separable functional forms are necessary. The modal decomposition method developed by Fergusson, Shellard and Liguori \cite{Fergusson:2006pr,Fergusson:2010dm} is well-suited for such purposes. So far it has been mainly applied to cases with less dramatic scale dependence, but should be able to be generalized to cases relevant to the signals studied here.

\medskip
\section*{Acknowledgments}
I would like to thank James Fergusson, Eugene Lim and Paul Shellard for helpful discussions. I am supported by the Stephen Hawking advanced fellowship.

\end{spacing}



\begin{thebibliography}{10}

\bibitem{Komatsu:2010fb}
  E.~Komatsu {\it et al.} [ WMAP Collaboration ],
  ``Seven-Year Wilkinson Microwave Anisotropy Probe (WMAP) Observations: Cosmological Interpretation,''
  Astrophys.\ J.\ Suppl.\  {\bf 192}, 18 (2011).
  [arXiv:1001.4538 [astro-ph.CO]].



\bibitem{Guth:1980zm}
  A.~H.~Guth,
  ``The Inflationary Universe: A Possible Solution To The Horizon And Flatness
  Problems,''
  Phys.\ Rev.\  D {\bf 23}, 347 (1981).

\bibitem{Linde:1981mu}
  A.~D.~Linde,
  ``A New Inflationary Universe Scenario: A Possible Solution Of The Horizon,
  Flatness, Homogeneity, Isotropy And Primordial Monopole Problems,''
  Phys.\ Lett.\  B {\bf 108}, 389 (1982).


\bibitem{Albrecht:1982wi}
  A.~J.~Albrecht and P.~J.~Steinhardt,
  ``Cosmology For Grand Unified Theories With Radiatively Induced Symmetry
  Breaking,''
  Phys.\ Rev.\ Lett.\  {\bf 48}, 1220 (1982).


\bibitem{Khoury:2001wf}
  J.~Khoury, B.~A.~Ovrut, P.~J.~Steinhardt, N.~Turok,
  ``The Ekpyrotic universe: Colliding branes and the origin of the hot big bang,''
  Phys.\ Rev.\  {\bf D64}, 123522 (2001).
  [hep-th/0103239].


\bibitem{Wands:1998yp}
  D.~Wands,
  ``Duality invariance of cosmological perturbation spectra,''
  Phys.\ Rev.\  {\bf D60}, 023507 (1999).
  [gr-qc/9809062].


\bibitem{Finelli:2001sr}
  F.~Finelli, R.~Brandenberger,
  ``On the generation of a scale invariant spectrum of adiabatic fluctuations in cosmological models with a contracting phase,''
  Phys.\ Rev.\  {\bf D65}, 103522 (2002).
  [hep-th/0112249].



\bibitem{Brandenberger:1988aj}
  R.~H.~Brandenberger, C.~Vafa,
 ``Superstrings in the Early Universe,''
  Nucl.\ Phys.\  {\bf B316}, 391 (1989).


\bibitem{Starobinsky:1979ty}
  A.~A.~Starobinsky,
  ``Relict Gravitation Radiation Spectrum and Initial State of the Universe. (In Russian),''
  JETP Lett.\  {\bf 30}, 682-685 (1979).



\bibitem{Kamionkowski:1996zd}
  M.~Kamionkowski, A.~Kosowsky, A.~Stebbins,
  ``A Probe of primordial gravity waves and vorticity,''
  Phys.\ Rev.\ Lett.\  {\bf 78}, 2058-2061 (1997).
  [arXiv:astro-ph/9609132 [astro-ph]].


\bibitem{Seljak:1996gy}
  U.~Seljak, M.~Zaldarriaga,
  ``Signature of gravity waves in polarization of the microwave background,''
  Phys.\ Rev.\ Lett.\  {\bf 78}, 2054-2057 (1997).
  [astro-ph/9609169].



\bibitem{Brandenberger:2006xi}
  R.~H.~Brandenberger, A.~Nayeri, S.~P.~Patil, C.~Vafa,
  ``Tensor Modes from a Primordial Hagedorn Phase of String Cosmology,''
  Phys.\ Rev.\ Lett.\  {\bf 98}, 231302 (2007).
  [hep-th/0604126].


\bibitem{Chen:2008wn}
  X.~Chen, R.~Easther, E.~A.~Lim,
  ``Generation and Characterization of Large Non-Gaussianities in Single Field Inflation,''
  JCAP {\bf 0804}, 010 (2008).
  [arXiv:0801.3295 [astro-ph]].


\bibitem{Chen:2011zf}
  X.~Chen,
  ``Primordial Features as Evidence for Inflation,''
  [arXiv:1104.1323 [hep-th]].


\bibitem{Chen:2010xk}
  X.~Chen,
  ``Primordial Non-Gaussianities from Inflation Models,''
  Adv.\ Astron.\  {\bf 2010}, 638979 (2010).
  [arXiv:1002.1416 [astro-ph.CO]].


\bibitem{Keisler:2011aw}
  R.~Keisler, C.~L.~Reichardt, K.~A.~Aird, B.~A.~Benson, L.~E.~Bleem, J.~E.~Carlstrom, C.~L.~Chang, H.~M.~Cho {\it et al.},
  ``A Measurement of the Damping Tail of the Cosmic Microwave Background Power Spectrum with the South Pole Telescope,''
  [arXiv:1105.3182 [astro-ph.CO]].

\bibitem{Hlozek:2011pc}
  R.~Hlozek, J.~Dunkley, G.~Addison, J.~W.~Appel, J.~R.~Bond, C.~S.~Carvalho, S.~Das, M.~Devlin {\it et al.},
  ``The Atacama Cosmology Telescope: a measurement of the primordial power spectrum,''
  [arXiv:1105.4887 [astro-ph.CO]].


\bibitem{Fergusson:2006pr}
  J.~R.~Fergusson, E.~P.~S.~Shellard,
  ``Primordial non-Gaussianity and the CMB bispectrum,''
  Phys.\ Rev.\  {\bf D76}, 083523 (2007).
  [astro-ph/0612713].

\bibitem{Fergusson:2010dm}
  J.~R.~Fergusson, M.~Liguori, E.~P.~S.~Shellard,
  ``The CMB Bispectrum,''
  [arXiv:1006.1642 [astro-ph.CO]].



\end{thebibliography}
\end{document}